Article Type: Review Article

# Synthetic Data for Robust AI Model Development in Regulated Enterprises

Aditi Godbole, *IEEE Senior Member, Bellevue, WA, USA*

*Abstract*—In today's business landscape, organizations need to find the right balance between using their customers' data ethically to power AI solutions and being compliant regarding data privacy and data usage regulations. In this paper, we discuss synthetic data as a possible solution to this dilemma. Synthetic data is simulated data that mimics the real data. We explore how organizations in heavily regulated industries, such as financial institutions or healthcare organizations, can leverage synthetic data to build robust AI solutions while staying compliant. We demonstrate that synthetic data offers two significant advantages by allowing AI models to learn from more diverse data and by helping organizations stay compliant against data privacy laws with the use of synthetic data instead of customer information. We discuss case studies to show how synthetic data can be effectively used in the finance and healthcare sector while discussing the challenges of using synthetic data and some ethical questions it raises. Our research finds that synthetic data could be a game-changer for AI in regulated industries. The potential can be realized when industry, academia, and regulators collaborate to build solutions. We aim to initiate discussions on the use of synthetic data to build ethical, responsible, and effective AI systems in regulated enterprise industries.

A financial institution trying to detect fraud or a healthcare institution looking for ways to identify diseases using X-rays needs a vast amount of data for these solutions. However, they must be careful about data privacy issues associated with the data they plan to use in building their AI-based solutions.

This is a problem many organizations are facing today, especially in industries with higher compliance and regulatory oversight, such as finance and healthcare. These companies wish to use AI and machine learning to build solutions that are efficient and scalable; however, they need to be cognizant of accidental data privacy and data security violations.

Synthetic data is a very promising solution for these situations, synthetic data is artificially generated data that mimics real-world data.

In this article, we will discuss methods of synthetic data creation, its role in Artificial Intelligence, and its potential to transform AI-driven innovations in regulated industries. We will explore how synthetic data can support organizations struggling to build AI systems that are simultaneously lawful, compliant, efficient, and scalable.

## Background

The regulatory landscape affecting AI and data usage in enterprises is complex and ever-changing. The variation in regulatory framework across industries and geographical regions makes it challenging for companies to adhere to regulations during AI model development. These regulations mainly revolve around transparency in data usage, fairness in AI decision-making processes, explainability of AI systems, and specific provisions for handling sensitive information.

Organizations face several key challenges in AI model development:

- **Data scarcity and quality:** Organizations often lack sufficient data and suffer from data quality issues, biases, and inconsistencies in datasets
- **Privacy concerns:** Organizations must ensure that their customer's personal information is not mishandled or misused, specifically in the healthcare and finance industries.
- **Regulatory compliance:** The continuously evolving regulatory environment makes it necessary for companies to be adaptable and vigilant.

Synthetic Data for AI development in regulated industries and Current Approaches Synthetic data generation is a potential solution to address these challenges. Synthetic data is artificially generated data that can mimic the properties of real data. Synthetic data allows organizations to comply with regulations while protecting user data[1]. By using synthetic data, companies can advance their applications quickly; for instance, Google's Waymo utilizes synthetic data to train its autonomous vehicles[2]. Synthetic data is widely adopted in the computer vision domain, where it is used for purposes such as augmenting training datasets and addressing class imbalance problems[3]. Synthetic data offers a promising avenue for companies to develop AI systems while adhering to strict privacy and regulatory requirements. It facilitates the creation of large, diverse datasets without compromising individual privacy or violating data protection laws[2].

## Synthetic Data Generation for Enterprise AI Development

As described in the previous section, synthetic data offers a promising solution for AI model development in regulated enterprise industries by allowing companies to adhere to data privacy standards and regulatory requirements. This section will discuss the practical aspects of using synthetic data for AI model development and evaluation strategies to ensure that the generated synthetic data is privacy-preserving and exhibits high fidelity compared to real data.

### Practical Implementation

To realize the benefits of synthetic data, we first need to understand how to generate high-quality synthetic data for our problem and how to evaluate it.

*Data Analysis and Preparation* In this step, the input dataset containing real data is analyzed to identify its statistical properties, distributions, and relationships between the data. Feature importance is another task performed to identify the most important and relevant features. Assessing biases or imbalances in the original data is one key step in this analysis.

*Synthetic Data Generation Techniques and Selection Methodology* The development of robust synthetic data solutions requires careful consideration of both technique selection and implementation methodologies. This section presents a discussion on key synthetic data generation approaches. We examine their characteristics, optimal application scenarios, and technical implementation requirements.

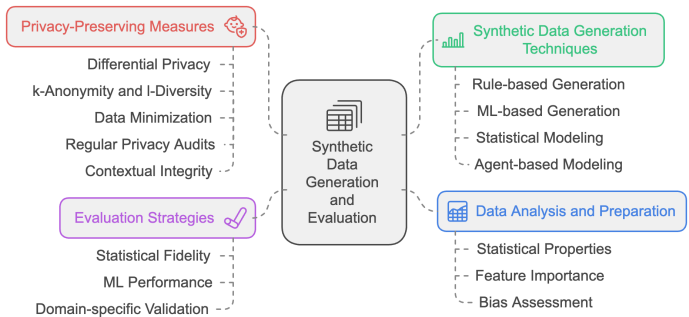

**FIGURE 1.** Synthetic Data Generation and Evaluation Process

*Rule-based generation*
Rule-based generation is one of the primary approaches in Synthetic data creation that addresses challenges in regulated enterprise environments. This methodology uses deterministic algorithms and expert-defined heuristics to generate synthetic data based on predefined rules and logic that reflect the characteristics and relationships found in real-world data. This method is particularly useful when domain experts have a clear understanding of the data structure and relationships, making it ideal for datasets that have well-defined structures and relationships. [4] The effectiveness of this approach depends on the ability to encode domain knowledge comprehensively during implementation.

Rule-based generation offers strong guarantees for data validity and regulatory compliance. However, it suffers from certain limitations. It may not capture complex relationships and patterns present in the real data, and managing large sets of interconnected rules becomes increasingly complex.

*Statistical Approaches*
In statistical approaches, synthetic data is generated by analyzing the original data and generating new data points that follow the same statistical distributions and correlations. The methods employ probability theory and statistical methods to analyze the original data and generate synthetic samples while maintaining the critical properties of the original dataset. [5] Distribution fitting, Copula-based methods, and Monte Carlo methods are some of the widely used statistical methods for synthetic data generation.

Distribution fitting techniques are advanced techniques for modeling univariate and multivariate distributions. These distributions are essential for capturing complex data relationships in financial risk modeling and healthcare outcomes analysis. These methods excel in scenarios requiring mathematical rigor and straightforward implementation but may struggle with complex pattern recognition.

Copulas are useful for capturing dependencies between multiple variables that have complex relationships and exhibit mixed data types, and copula-based methods and vine copulas can be utilized to preserve complex multivariate relationships in synthetic data [6], [7]. In financial data or in sensor readings where variables have complex correlations, copulas allow us to maintain these correlations while generating synthetic data. These methods may face scalability challenges with very large datasets.

Monte Carlo methods are widely applied in risk assessment and scenario generation by generating synthetic samples through repeated random sampling. [8]

*ML-based generation*

Machine learning approaches, particularly Generative Adversarial Networks (GANs) [9] and Variational Autoencoders (VAEs),[10] offer powerful capabilities for capturing complex data patterns while maintaining privacy guarantees. It can create data that closely mimics the statistical properties and patterns of the original dataset[11].

GANs employ an adversarial framework comprising two neural networks: a generator creating synthetic instances and a discriminator evaluating their authenticity. This architecture enables the generation of highly realistic synthetic data through iterative optimization. In regulated enterprise environments, GANs have demonstrated particular efficacy in generating synthetic medical imaging data and financial transaction patterns while preserving statistical relationships and privacy constraints [12]. GANs demonstrate particular efficacy in generating high-dimensional data, though they require substantial computational resources and expertise in managing training stability.

VAEs provide an alternative approach through probabilistic encoding-decoding architectures. By learning compact latent representations of input data, VAEs can generate diverse synthetic samples while maintaining essential statistical properties. This methodology proves especially valuable for structured data generation in healthcare and financial applications, where maintaining complex variable relationships is crucial [3]. VAEs offer a balanced alternative, providing robust performance for structured data while maintaining reasonable computational requirements.

*Agent-based modeling*

Agent-based modeling represents a distinct paradigm in synthetic data generation, This approach is particularly useful for modeling complex systems with many interacting parts[13]. Agentbased modeling simulates the actions and interactions of autonomous agents within a system to generate synthetic data.

The strength of this methodology lies in its ability to generate synthetic data that can closely reflect the complex system dynamics and maintain realistic causal relationships in multi-agent systems while capturing temporal changes and interaction patterns

*Hybrid approaches*

In practical applications, synthetic data generation often involves integrating multiple methodologies. A common approach employs statistical techniques to establish foundational data structures. Machine learning algorithms are then used to introduce intricate patterns. Rule-based validation is subsequently applied to ensure compliance with specific business constraints. [15]

This hybrid methodology combines the strengths of each technique. Machine learning provides flexibility to capture complex patterns. Rule-based systems offer precision and control. Statistical methods ensure that the generated data adheres to rigorous standards of validity. The combination of these approaches facilitates the creation of synthetic datasets that not only mirror the statistical properties of real data but also comply with domain-specific rules and constraints. This ensures that the generated data is both realistic and relevant for practical applications.

## Technical Implementation Considerations

The implementation of synthetic data generation solutions requires careful consideration of multiple technical factors and decision criteria. When selecting appropriate synthetic data generation techniques, organizations must evaluate their specific requirements across several critical dimensions: data characteristics, computational resources, and intended application scenarios. Data type serves as the primary criterion in technique selection, as it significantly influences the viability of different generation approaches. One of the key requirements while choosing the right synthetic data generation technique is that the synthetic data should maintain the statistical properties of the original data while preserving the anonymity of individual records.

Resource considerations play an important role in implementation decisions. Statistical approaches typically demand minimal computational resources, making them suitable for rapid prototyping or scenarios with limited infrastructure. Conversely, deep learning methods like GANs require significant computational power and specialized expertise, which makes careful evaluation of available resources against desired outcomes a necessity.

Dataset size represents another critical factor influencing technique selection. While some methods require substantial training data to produce highquality synthetic data, others can perform effectively with smaller datasets. Statistical approaches and copula-based methods often perform adequately with modest dataset sizes, whereas deep learning approaches typically require larger training sets to achieve optimal results.

The selection of synthetic data generation techniques requires a rigorous evaluation of multiple factors, with a primary focus on maintaining statistical fidelity while ensuring robust privacy protection. [16] Organizations must carefully evaluate the trade-offs between data utility, computational requirements, and privacy guarantees, considering the specific context and requirements of each implementation scenario. This comprehensive evaluation framework is essential for determining the most appropriate methodology that can effectively balance these competing demands.

## Privacy-preserving and Regulatory Compliance Measures

Synthetic data is not inherently privacy-preserving. It is artificially generated data; it does not directly contain real individuals' information. However, it can still suffer from privacy risks since it may be possible to infer information about real individuals from synthetic data, as the synthetic data very closely mimics the original data. Machine learning models used for synthetic data generation could accidentally memorize and reproduce the sensitive information from training data.

To make synthetic data privacy-preserving, additional privacy-preserving techniques must be implemented:

- **Differential Privacy:** Incorporate differential privacy techniques into the data generation process to add controlled noise and provide mathematical privacy guarantees. This methodology ensures that the presence or absence of any individual record cannot be reliably inferred from the synthetic dataset.[17]
- **k-Anonymity, l-Diversity, and t-Closeness:** Ensure that synthetic data adheres to k-anonymity which means each record is indistinguishable from at least k-1 others[18], in l-diversity sensitive attributes have at least one well-represented values[19] and t-closeness ensures that the distribution of a sensitive attribute within a generalization of a quasi-identifier is close to the distribution of the sensitive attribute in the entire dataset[20]. Each of these techniques is a progression in privacy protection, where each technique addresses limitations from the previous one. These techniques are used in combination or as part of a more complex privacy-preserving approach.

In regulated industries, these privacy-preserving techniques must align with specific regulatory frameworks. In the United States, financial institutions must adhere to model risk management requirements under SR 11-7, necessitating comprehensive documentation of synthetic data generation processes and their impact on model performance. Healthcare organizations must ensure HIPAA compliance, requiring rigorous validation of synthetic data generation processes and preservation of clinical utility while maintaining patient privacy [21].

In regulated industries, these privacy-preserving techniques must align with specific regulatory frameworks. In the United States, financial institutions must adhere to model risk management requirements under SR 11-7, necessitating comprehensive documentation of synthetic data generation processes and their impact on model performance. Healthcare organizations must ensure HIPAA compliance, requiring rigorous validation of synthetic data generation processes and preservation of clinical utility while maintaining patient privacy.

- **Data Minimization:** Generate only the necessary attributes and records needed for the specific use case, reducing the risk of unnecessary information disclosure[22].
- **Regular Privacy Audits:** Conduct frequent privacy risk assessments on the synthetic data to identify and mitigate potential vulnerabilities.
- **Contextual Integrity:** Consider the context in which the synthetic data will be used and ensure that it does not violate individuals' privacy expectations within that context.

Documentation requirements for regulatory compliance encompass detailed records of generation processes, validation test results, and privacy impact as-sessments. Organizations must maintain comprehensive audit trails that demonstrate ongoing compliance with privacy preservation measures and regulatory requirements. This documentation serves both as evidence of compliance and as a foundation for continuous improvement of privacy protection mechanisms.

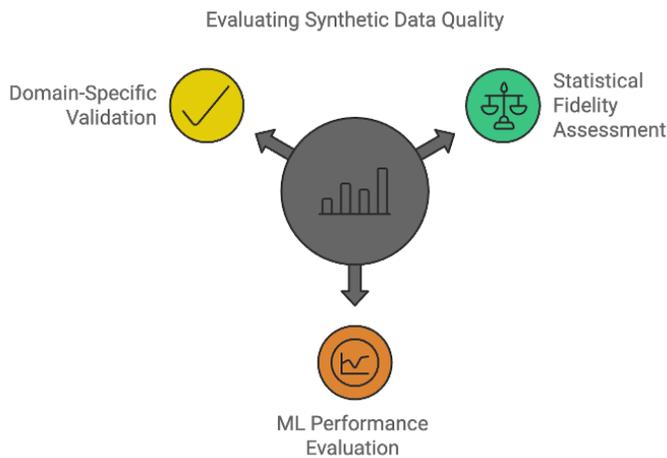

**FIGURE 2.** Synthetic Data Generation Quality Evaluation Process

By implementing these measures, organizations can improve the privacy-preserving properties of synthetic data and follow regulatory compliance standards. Privacy preservation and regulatory compliance are an ongoing process that requires continuous evaluation and adjustment.

## Scaling Considerations

The scalability and performance issues associated with generating large volumes of synthetic data can be resolved using distributed computing frameworks for data generation. The use of efficient storage and retrieval techniques must be used for the synthetic datasets and the system should have the ability to keep track of data distribution changes to update synthetic data continuously to follow the original data distribution changes.

## Evaluation Strategies

It is important to evaluate the quality of synthetic data generated to ensure it follows the original data distribution and provides robust performance for the trained machine model.

*Statistical fidelity assessment*
The fidelity of synthetic data can be measured through various statistical measures such as KL divergence and maximum mean discrepancy that can quantify the similarity between real and synthetic data distributions[12].

Another technique of evaluating the quality of synthetic data is using the concepts from adversarial machine learning, where a model is employed to evaluate whether it can distinguish between the real and synthetic data, which helps us understand the quality of synthetic data[23].

*ML performance*
One method of evaluating the generalization capability of generated synthetic data is to assess how well the model trained on synthetic data performs on a real-world test data set[12].

*Domain-specific validation*
Developing and applying domain-specific evaluation metrics can help ensure that models trained on synthetic data capture the nuances and complexities of the target domain [24]. For instance, in financial risk modeling, metrics such as Value at Risk (VaR) and Expected Shortfall (ES) have been used to validate models trained on synthetic market data [25].

## Case Studies in Regulated Industries

We will examine two case studies that demonstrate different aspects of synthetic data implementation in regulated industries. The first case study presents quantitative results for a financial services implementation, showing concrete evidence of synthetic data's effectiveness in fraud detection. The second case study outlines a theoretical framework for healthcare diagnostics, illustrating how organizations can systematically plan synthetic data implementation in domains with stringent privacy requirements.

The financial services industry faces significant challenges in developing robust fraud detection models due to the sensitive nature of financial data and the rarity of actual fraud events. Synthetic data can be used to tackle both these challenges.

*Problem statement:*
A major bank needs to improve its fraud detection capabilities. Due to the regulations and privacy concerns, it cannot use the transaction data, and the data also suffers from a scarcity of fraud examples.

*Solution:*
The bank can generate synthetic transaction data using synthetic data generation techniques such as GAN and include a range of fraud examples. This synthetic data will maintain the properties of real transaction data while not containing any actual customer information. It can be used to train a model that can predict fraudulent transactions. While generating the synthetic data, typically, a GAN model will be trained on anonymized historical transaction data, and differential privacy techniques will be applied to ensure individual privacy.

Our research demonstrates how synthetic data can effectively address both challenges - privacy preservation and imbalanced dataset, through a detailed implementation study. In a comprehensive fraud detection case study, we worked with a credit card transaction dataset comprising 284,807 transactions, of which only 0.17% (492 transactions) represented fraudulent activities. This extreme imbalance, combined with the sensitive nature of transaction data, presented an ideal scenario for synthetic data application. We implemented a Conditional Generative Adversarial Network (CGAN) architecture specifically designed for generating synthetic financial transaction data. The generator network processed a 128-dimensional noise vector combined with class labels through multiple dense layers with LeakyReLU activations and batch normalization, ultimately producing synthetic transactions matching the 29-dimensional feature space of real transactions. The discriminator network evaluated these generated transactions alongside real ones, helping refine the generation process through adversarial training.

Training the CGAN over 1000 epochs revealed systematic improvement in model performance. The discriminator loss decreased from 1.1534 to 0.4440, indicating increasing ability to distinguish real from synthetic data. Meanwhile, the generator loss increased from 0.8335 to 1.5519, reflecting the growing complexity of the adversarial game as the generator worked harder to create more convincing synthetic samples.

To validate the effectiveness of our synthetic data, we conducted a comparative analysis using two Random Forest classifiers: one trained on real data and another on synthetic data. Both models were evaluated on the same real-world test set. The model trained on real data achieved an AUC-ROC score of 0.96, while the model trained on synthetic data achieved a comparable AUC-ROC of 0.93, demonstrating that synthetic data preserved the discriminative patterns crucial for fraud detection.

The confusion matrices revealed that the real-datatrained model achieved a recall rate of 76detection, while the synthetic-trained model achieved a recall rate of 22synthetic-trained model maintained some ability to detect fraud despite no real transaction data being stored or exposed during model deployment. This approach effectively addresses privacy concerns, though further improvements in recall rates are needed to enhance the robustness of fraud detection using synthetic data.

The synthetic data successfully captured the multidimensional relationships between transaction features while introducing sufficient variation to prevent privacy leakage through memorization. Building on the insights from the financial sector implementation, we next examine how synthetic data approaches can be adapted for healthcare applications. While the financial sector case study demonstrated immediate practical results, our healthcare case study presents a comprehensive framework for future implementation, addressing the distinct regulatory and technical requirements of medical applications. In the healthcare domain, companies must comply with strict privacy regulations such as HIPAA. It is also challenging to gain access to diverse patient data. Synthetic data, along with privacypreserving techniques, provides robust solutions.

Building on the insights from the financial sector implementation, we next examine how synthetic data approaches can be adapted for healthcare applications. While the financial sector case study demonstrated immediate practical results, our healthcare case study presents a comprehensive framework for future implementation, addressing the distinct regulatory and technical requirements of medical applications. In the healthcare domain, companies must comply with strict privacy regulations such as HIPAA. It is also challenging to gain access to diverse patient data. Synthetic data, along with privacy-preserving techniques, provides robust solutions.

*Problem statement:*
A healthcare institution is developing a diagnostic tool to identify and diagnose a rare disease associated with a respiratory condition. Due to privacy regulations, they do not have access to real patient data. Also, the rare nature of the disease makes it more challenging to train a robust model.

*Solution:*
The healthcare institution can apply a federated learning approach to the data from several associated hospitals. Model training will be done on synthetic data generated using a GAN model. With this approach, healthcare institutions will be able to train a privacy-preserving model to diagnose rare diseases. With this synthetic data, the healthcare institution will be able to train a robust model for rare disease detection, and by using federated learning, the healthcare institution is able to make sure that no real patient data leaves the hospitals' secure environments.

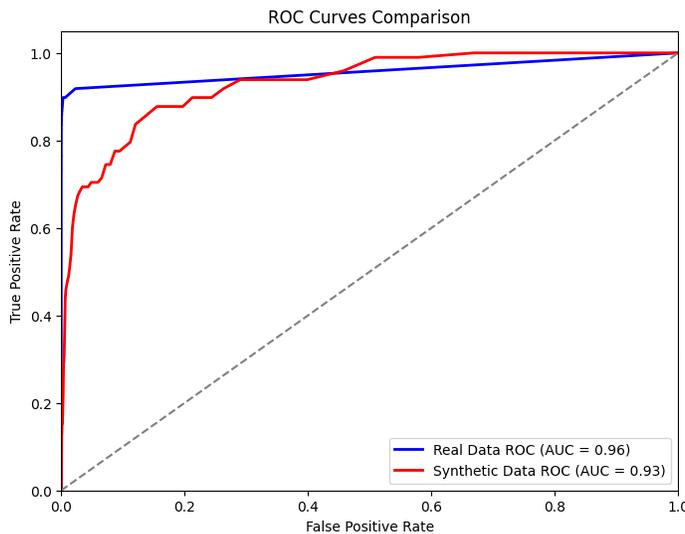

**FIGURE 3.** Synthetic Data Generation and Evaluation Process

The proposed implementation framework begins with data collection and preparation phases. The institution would establish data sharing agreements with multiple affiliated hospitals to collect anonymized respiratory diagnostic data. The dataset structure would encompass comprehensive patient diagnostics, including vital signs, laboratory results, imaging metrics, and validated patient outcomes. The presence or absence of a rare respiratory condition will be the target variable.

For synthetic data generation, the framework employs a GAN architecture incorporating differential privacy guarantees. The generator component consists of a five-layer neural network designed to process a 256-dimensional noise vector, while the discriminator utilizes a four-layer network architecture with batch normalization. To ensure HIPAA compliance, the system maintains a privacy budget of 1.0, striking a balance between data utility and privacy preservation.

The evaluation framework encompasses multiple complementary approaches to validate both the quality of synthetic data and its utility for rare disease diagnosis. Statistical similarity between real and synthetic distributions would be measured using KL divergence metrics, providing quantitative validation of the synthetic data's fidelity. Clinical validity assessment would involve domain experts reviewing generated cases to ensure medical coherence and plausibility.

For model performance evaluation, the framework incorporates comprehensive metrics to assess diagnostic capability. This includes classification accuracy measurement, sensitivity and specificity analysis specifically focused on rare disease detection, and area under the ROC curve to evaluate overall model discrimination ability. Privacy protection would be validated through resistance testing against membership inference and attribute inference attacks, ensuring robust protection of patient privacy.

The proposed federated learning integration enables hospitals to contribute to model training without exposing sensitive patient data. Each participating institution would maintain local model training using their proprietary patient data while only sharing model parameters rather than actual patient records. This approach ensures HIPAA compliance while leveraging diverse patient populations for improved model robustness.

To ensure ongoing compliance and effectiveness, the framework includes continuous monitoring and validation protocols. Regular assessment of synthetic data quality, model performance, and privacy preservation metrics would guide iterative improvements to the synthetic data generation process. This systematic approach enables healthcare institutions to advance their diagnostic capabilities while maintaining strict adherence to privacy regulations and ethical guidelines.

Through this theoretical framework, we demonstrate how healthcare organizations can systematically approach the challenge of developing AI diagnostic tools using synthetic data, even for rare conditions where real patient data is limited. The structured evaluation methodology provides a roadmap for validating both the technical effectiveness and regulatory compliance of synthetic data solutions in healthcare applications.

## Challenges and Limitations

Synthetic data offers significant potential for developing robust AI models in regulated industries. However, there are challenges and limitations that we should discuss. It is essential that synthetic data accurately represents the complexities and nuances of real-world data when building robust AI models. In domains like rare disease diagnosis or fraud detection, synthetic data should be able to represent infrequent but important scenarios.

The synthetic data should preserve the complex relationships between variables in the original realworld data. The loss of this information can lead to inaccurate model predictions and compromised integrity of any decision-making processes based on the models trained using synthetic data. The need for robust metrics to evaluate the quality and fidelity of synthetic data is still an active area of research for getting an understanding of synthetic data quality. Propensity score matching and statistical similarity measures are used to provide insight into synthetic data quality.

In regulated industries, organizations using synthetic data for AI model development must develop rigorous validation processes to show that models trained on synthetic data perform reliably on real-world data. Regulators may require detailed synthetic data generation process documentation to ensure integrity and fairness.

Model interpretability is crucial for regulatory compliance and stakeholder trust in many regulated industries. Techniques for model interpretability may need to be adapted to account for the use of synthetic training data for explaining AI models trained on synthetic data.

## Future Directions

As we look ahead, several key areas warrant further exploration in the field of synthetic data for AI in regulated industries:

- **Advanced Generation Techniques:** Developing more sophisticated algorithms to improve the fidelity of synthetic data, especially for complex, multidimensional datasets typical in finance and healthcare.
- **Domain-Specific Solutions:** Creating tailored synthetic data solutions for different regulated industries, including specialized evaluation metrics and validation processes.
- **Scalability and Efficiency:** Exploring distributed computing frameworks and optimized algorithms capable of generating high-quality synthetic data at scale.
- **Regulatory Frameworks:** Collaborating with researchers, industry practitioners, and regulators to develop standardized frameworks for synthetic data use in AI development.
- **Explainable AI:** Advancing techniques for model interpretability specifically for AI models trained on synthetic data, crucial for regulatory compliance in many industries.
- **Federated Learning Integration:** Investigating synergies between federated learning and synthetic data generation for enhanced privacy-preserving AI development.
- **Ethical Considerations:** Developing methods to mitigate potential biases introduced or amplified through synthetic data generation, ensuring fair representation of diverse populations.
- **Real-time Generation:** Exploring techniques for real-time or near-real-time synthetic data generation to enable adaptive AI systems in dynamic environments.
- **Cross-Industry Applications:** Adapting synthetic data techniques across different regulated industries to accelerate progress and foster innovation.

By advancing these areas, we can further unlock the potential of synthetic data in AI development for regulated industries, moving towards a future where organizations can harness AI's power while maintaining the highest standards of privacy, security, and regulatory compliance.

## Conclusion

In this paper, we have explored the potential of synthetic data as a solution for AI development in regulated industries. Organizations in regulated sectors like finance and healthcare face the dual challenges of using customer data for AI solutions and adhering to stringent privacy regulations; synthetic data is a promising solution

We have discussed various methods of synthetic data generation, including rule-based, ML-based, statistical modeling, and agent-based approaches. Each technique offers unique advantages depending on the specific use case and data characteristics. We have also emphasized the critical importance of privacy-preserving measures and regulatory compliance when generating and using synthetic data.

Our case studies in the financial and healthcare sectors demonstrate the practical applications of synthetic data in overcoming real-world challenges. From enhancing fraud detection capabilities in banking to enabling rare disease diagnosis in healthcare, synthetic data proves its versatility and effectiveness.

However, it is important to acknowledge the challenges and limitations associated with synthetic data. Ensuring the fidelity of synthetic data, preserving complex relationships between variables, and developing robust evaluation metrics remain active areas of research. Moreover, the need for rigorous validation processes and regulatory acceptance poses additional hurdles. In regulated industries, synthetic data can provide great benefits in building ethical, responsible, and effective AI systems.

In conclusion, synthetic data offers a path to innovation that balances the need for data-driven insights with the imperative of privacy protection. As AI continues to transform regulated industries, synthetic data will undoubtedly play a crucial role in shaping the future of ethical and compliant AI development.

Aditi Godbole Aditi Godbole is a Senior Data Scientist with over 11 years of experience in machine learning and artificial intelligence. She specializes in natural language processing, supervised learning, and generative AI techniques. Her work focuses on developing enterprise-scale solutions to address complex business challenges. Aditi Godbole has led numerous largescale software projects and contributes to shaping AI and ML strategies in her current role. She holds patents in the field and is actively involved in mentoring and knowledge sharing within the ML and Data Science community.